\documentclass[english,aps,pra,reprint,showpacs,amsfonts,amsmath,amssymb,nofootinbib]{revtex4-1}
\usepackage[english]{babel}
\usepackage{amsfonts}
\usepackage{amsmath}
\usepackage{amssymb}
\usepackage{graphicx}
\usepackage{enumerate}
\usepackage{color}

\begin{document}

\def\arXiv#1#2#3#4{{#1} #2 #3 {\it Preprint} #4}
\def\Book#1#2#3#4#5{{#1} {\it #3} (#4, #5, #2).}
\def\Bookwd#1#2#3#4#5{{#1} {\it #3} (#4, #5, #2)}
\def\Journal#1#2#3#4#5#6#7{#1, #4 \textbf{#5}, #6 (#2).}
\def\JournalE#1#2#3#4#5#6{#1, #4 \textbf{#5}, #6 (#2).}
\def\eref#1{(\ref{#1})}

\newcommand{\dd}{\mbox{d}}
\newcommand{\EE}{\mathbb{E}}
\newcommand{\NN}{\mathbb{N}}
\newcommand{\PP}{\mathbb{P}}
\newcommand{\RR}{\mathbb{R}}
\newcommand{\TT}{\mathbb{T}}
\newcommand{\ZZ}{\mathbb{Z}}
\newcommand{\uu}{\mathbf{1}}
\newcommand{\HH}{\mathcal{H}}

\title{
Invariance in quantum walks with time-dependent coin operators}
\author{Miquel Montero}
\email{miquel.montero@ub.edu}
\affiliation{Departament de F\'{\i}sica Fonamental, Universitat de Barcelona (UB), Mart\'{\i} i Franqu\`es 1, E-08028 Barcelona, Spain}
\date{\today}


\begin{abstract}
In this paper we unveil some features of a discrete-time quantum walk on the line whose coin depends on the temporal variable. After considering the most general form of the unitary coin operator, we focus on the role played by the two phase factors that one can incorporate there, and show how both terms influence the evolution of the system. A closer analysis reveals that the probabilistic properties of the motion of the walker remain unaltered when the update rule of these phases is chosen adequately. This invariance is based on a symmetry with consequences not yet fully explored.
\end{abstract}
\maketitle

\section{Introduction}

The quantum walk (QW)~\cite{ADZ93,TM02,NK03,JK03,VA12} was primarily devised as the quantum-mechanical version of the classical random walk, the stochastic process modeling the trajectory of a particle that at each time step moves, either leftward or rightward, a fixed distance in accordance with the outcome of a coin toss. In the quantum version, the coin is related to some intrinsic degree of freedom in the system with a quantum nature and two exclusive values, a qubit: e.g., the spin, the polarization or the chirality.

It soon became clear that, beyond the formal resemblance, random and quantum walks show very different properties~\cite{CFG03}, being perhaps the most striking of them the ability of QWs to spread over the line, not as a diffusive process, but linearly in time~\cite{ABNVW01}. This speed-up makes feasible the development of quantum algorithms that can solve problems in a more efficient way than their classic counterparts~\cite{PS97,FG98}. In particular, QWs are promising tools in the optimal resolution of search problems~\cite{SKW03,AMB10,MNRS11}. Nowadays, QWs have exceeded the scope of quantum computation and attracted the attention of many researchers from other fields as, for example, game theory~\cite{FAJ04,BFT08,CB11,RH11}.

In fact, Refs.~\cite{FAJ04,BFT08,CB11,RH11} are paradigmatic instances that show how, fruit of this broad interest, diverse extensions of the discrete-time QW on the line have been considered in the past. Most of these variations are related with the properties of the unitary coin operator, cornerstone of the new physical features of the system. Thus, one can find in the literature QWs whose evolution depends on more than one coin~\cite{BCA03a,TFMK03,VBBB05}, QWs that suffer from decoherence~\cite{BCA03b,KT03}, or QWs driven by inhomogeneous, site-dependent coins~\cite{WLKGB04,RASAD05,SK10,KLS13,ZXT14,XQTS14}, just to name a few.

These all are examples of modified QWs in which the coin changes with time in an implicit way. But there are also precedents where the temporal inhomogeneity of the QW is made absolutely explicit: in the form of a recursive rule for the coin selection, as in the so-called Fibonacci QWs~\cite{RMM04,AR09a}, though a given function that determines the value of the coin parameters~\cite{AR09b,RS14,BNPRS06}, or by means of a random process that controls the coin properties~\cite{AVWW11}.

From all of these previous works, it is in Ref.~\cite{BNPRS06} where one can find more similarities with respect to the path we are going to follow in this paper. Here we consider the evolution of a discrete-time QW on the line with a general, time-dependent coin. The generality in our analysis stems from the inclusion of the two phase factors that the unitary coin operator can incorporate, a caution that may become superfluous depending on the circumstances~\cite{TFMK03}. In the present case, this generalization is essential: as the authors of Ref.~\cite{BNPRS06} rightly pointed out, these phase factors can be used to induce new features in the QW (like quasiperiodic behavior or dynamic localization) but also as a control mechanism for compensating externally-induced decoherence. Our contribution shares these two qualities as well: we show how the space evolution of the system remains unaltered from the probabilistic point of view if the phase factors are well tuned. Thus, this nontrivial invariance is added to other known symmetries of the problem~\cite{CSB07}.

The paper is organized as follows. In Sec.~\ref{Sec_Process} we review the formalism used in the construction of the discrete-time quantum walk on the line with a time-dependent coin operator. In Sec.~\ref{Sec_Initial}  we provide explicit expressions for the initial stages of the evolution of the quantum state of the system. We analyze the mathematical structure of these formulas in Sec.~\ref{Sec_Changing_phases}, and infer that invariant behavior can be obtained with the proper selection of the values of the parameters. We prove the general validity of this conjecture in Sec.~\ref{Sec_Invariance}, and discuss the role of the two key magnitudes that appear along our study. The paper ends with Sec.~\ref{Sec_Conclusion} where conclusions are drawn.

\section{QW with a time-dependent coin operator}
\label{Sec_Process}
We begin by introducing the building blocks of the time-dependent quantum walk on the line.  We denote by $\HH_P$ the Hilbert space of discrete particle positions in one dimension, spanned by the basis $\left\{| n\rangle : n \in \ZZ\right\}$, and by $\HH_C$ the Hilbert space of the coin states, spanned by the basis $\left\{|+\rangle, |-\rangle\right\}$. In its most ubiquitous form, the discrete-time, discrete-space quantum walk on the Hilbert space $\HH\equiv\HH_C\otimes \HH_P$ is the result of the action of the evolution operator $\hat{T}$, $\hat{T}\equiv \hat{S}\, \hat{U}$, which consists of some unitary operator $\hat{U}$ that affects only the internal degree of freedom, the throw of the quantum dice, and a shift operator $\hat{S}$ that {\it moves\/} the walker depending on the respective coin state:
\begin{equation}
\hat{S} \left(|\pm\rangle\otimes| n\rangle \right)= |\pm\rangle\otimes| {n\pm 1}\rangle.
\end{equation}

In our scheme, however, we break the temporal homogeneity of the process by assuming that the coin operator changes with time, $\hat{T}_t$, i.e., $\hat{T}_t\equiv \hat{S}\, \hat{U}_t$. The most general form that operator $\hat{U}_t$ can take according to its unitary nature reads~\cite{CSL08}:
\begin{eqnarray}
\hat{U}_t
&\equiv& \big[e^{i \alpha_t} \cos \theta_t |+\rangle  \langle +| +e^{-i \beta_t} \sin \theta_t |+\rangle  \langle -| \nonumber \\
&+& e^{i \beta_t} \sin \theta_t  |-\rangle  \langle +| - e^{-i \alpha_t} \cos \theta_t  |-\rangle  \langle -|\big]\otimes\hat{I}_P,
\label{U_coin_gen}
\end{eqnarray}
where $\hat{I}_P$ is the identity operator defined in the position space $\HH_P$, and $\alpha_t$, $\beta_t$ and $\theta_t$ are real magnitudes. In fact, as we consider that the time increases in discrete steps, one can always choose the time units so that the time variable $t$ is just an integer index. Then the state of the system at a later time, $|\psi\rangle_t$, is recovered after the application of $\hat{T}_t$ to the previous state:   
\begin{equation}
|\psi\rangle_t =\hat{T}_t|\psi\rangle_{t-1},
\label{evol_t}
\end{equation}
and the evolution of the system is fully determined once $|\psi\rangle_{0}\equiv|\psi\rangle_{t=0}$ is set. Here we assume that, at the beginning, the quantum walker is located at the origin but that the initial coin state is a general superposition:
\begin{equation}
|\psi\rangle_{0} =\left(\cos \eta |+\rangle + e^{i \gamma}\sin \eta   |-\rangle\right) \otimes | 0\rangle.
\label{psi_zero_gen}
\end{equation}

Let us finally introduce two probabilistic magnitudes which will appear repeatedly along this paper. The first one is the probability mass function (PMF) of the process, $\rho(n,t)$, the probability that the walker is in a particular position $n$ at a given time $t$:
\begin{equation}
\rho(n,t)=\langle \psi|\hat{N}_n|\psi\rangle_t,
\end{equation}
where
\begin{equation}
\hat{N}_n\equiv \hat{I}_C\otimes  | {n}\rangle\langle n|,
\label{N_def}
\end{equation}
and $\hat{I}_C$ is the identity operator of the coin space $\HH_C$. The second one is the probability of obtaining the {\it plus\/} or {\it minus\/} value when measuring the coin state of the walker:
\begin{equation}
P_{\pm}(t)\equiv\langle \psi|\hat{Q}_{\pm}|\psi\rangle_t,
\end{equation}
with
\begin{equation}
\hat{Q}_{\pm}\equiv  | \pm\rangle\langle \pm| \otimes\hat{I}_P.
\label{S_def}
\end{equation}

\section{Initial evolution}
\label{Sec_Initial}

We are now ready to determine the state of  the system $|\psi\rangle_{t}$, for any value of $t$, by means of Eqs.~\eqref{evol_t} and~\eqref{psi_zero_gen}. We might give specific values to $\eta$ and $\gamma$, and a rule for $\alpha_t$, $\beta_t$ and $\theta_t$, and thus $|\psi\rangle_{t}$ could be computed numerically. Instead of proceeding in this way, we are going to introduce the explicit expressions of $|\psi\rangle_{t}$ for the lowest values of $t$. As we will see, beyond the obvious limitations, this approach is very illustrative.

Thus, after the first time step the system is in the state given by
\begin{eqnarray}
|\psi\rangle_{1} &=& e^{i\alpha_1}\left[\cos \eta\cos \theta_1+e^{-i \varphi}\sin \eta\sin\theta_1 \right]  |+\rangle\otimes | 1\rangle \nonumber\\
&+&e^{i\beta_1}\left[\cos \eta\sin \theta_1-e^{-i \varphi}\sin \eta\cos\theta_1 \right]  |-\rangle\otimes | {-1}\rangle, \nonumber\\
&=&e^{i\alpha_1} a \, |+\rangle\otimes | 1\rangle+e^{i\beta_1} b\,  |-\rangle\otimes | {-1}\rangle,
\label{psi_one}
\end{eqnarray}
where
\begin{eqnarray}
a&\equiv&\cos \eta\cos \theta_1+e^{-i \varphi}\sin \eta\sin\theta_1, \\
b&\equiv&\cos \eta\sin \theta_1-e^{-i \varphi}\sin \eta\cos\theta_1,
\end{eqnarray}
with $\left|a\right|^2+\left|b\right|^2=1$, and
\begin{equation}
\varphi\equiv\alpha_1+\beta_1-\gamma.
\label{phi_def}
\end{equation}

In this case, only two sites in the line are accessible to the walker, i.e., $n=\pm1$,
\begin{eqnarray}
\rho(1,1)&=&\frac{1}{2}\left(1+\cos 2\eta\cos 2\theta_1 +\sin 2\eta\sin2\theta_1\cos \varphi\right)\nonumber\\
&=&\left|a\right|^2,\\
\rho(-1,1)&=&\frac{1}{2}\left(1-\cos 2\eta\cos 2\theta_1 -\sin 2\eta\sin2\theta_1\cos \varphi\right)\nonumber\\
&=&\left|b\right|^2,
\label{rho_one}
\end{eqnarray}
which is nothing but the consequence of a well-known property of quantum walks of this kind: $\rho(n,t)=0$ if 
$n\neq t-2 m$, $m\in\{0,\ldots,t\}$. The same expressions are obtained for the probabilities of the two possible outputs after a measure of the coin state of the system: 
\begin{equation}
P_{\pm}(t=1)=\rho(\pm1,1).
\end{equation}

The state of the system after the second time step is
\begin{eqnarray}
|\psi\rangle_{2} &=&e^{i(\alpha_1+\alpha_2)}\cos\theta_2 \,a |+\rangle\otimes | {2}\rangle \nonumber\\
&+&\sin\theta_2\left[e^{i(\beta_1 - \beta_2)} b |+\rangle +e^{i (\alpha_1+\beta_2)}\,a  |-\rangle\right] \otimes | {0}\rangle\nonumber\\
&-&e^{i(\beta_1-\alpha_2)}\cos\theta_2\,b  |-\rangle\otimes | {-2}\rangle,
\label{psi_two}
\end{eqnarray}
and depends on  $\alpha_2$, $\beta_2$ and $\theta_2$. Nonetheless, the PMF is independent of  $\alpha_2$ and $\beta_2$,
\begin{eqnarray}
\rho(2,2)&=&\cos^2\theta_2\, \left|a\right|^2,\\
\rho(0,2)&=&\sin^2\theta_2,\\
\rho(-2,2)&=&\cos^2\theta_2\, \left|b\right|^2;
\label{rho_two}
\end{eqnarray}
and the same applies for the probabilities of measuring any of the two possible values of the coin: 
\begin{equation}
P_{\pm}(t=2)=\cos^2\theta_2\,\rho(\pm1,1)+\sin^2\theta_2\,\rho(\mp1,1).
\end{equation}

This fact could lead to the impression that $\alpha_t$ and $\beta_t$ bear no physical information and thus to concluding that one can freely set $\alpha_t=\beta_t=0$. Note that with this convention the values of $\rho(\pm1,1)$ can be kept unchanged with a suitable redefinition of $\gamma$. This is one of the most standard approaches used in the literature when $\hat{U}$ does not change~\cite{TFMK03}. As we will see, one has to proceed with more caution when the coin operator evolves~\cite{CSB07}.

After the third iteration, the state of the system is 
\begin{eqnarray}
|\psi\rangle_{3} &=&e^{i(\alpha_1+\alpha_2+\alpha_3)}\cos\theta_3\cos\theta_2\,a  |+\rangle\otimes | {3}\rangle \nonumber\\
&+&\Big\{\Big[e^{i(\beta_1-\beta_2+\alpha_3)}\cos\theta_3\sin\theta_2\,b\nonumber\\
&+&e^{i(\alpha_1+\beta_2-\beta_3)}\sin\theta_3\sin\theta_2 \,a\Big]|+\rangle \nonumber\\
&+&e^{i(\alpha_1+\alpha_2+\beta_3)}\sin\theta_3\cos\theta_2\, a |-\rangle \Big\}\otimes| {1}\rangle \nonumber\\
&-&\Big\{e^{i(\beta_1-\alpha_2-\beta_3)}\sin \theta_3\cos\theta_2\,b|+\rangle \nonumber\\
&+&\Big[e^{i(\alpha_1+\beta_2-\alpha_3)}\cos\theta_3\sin\theta_2\,a\nonumber\\
&-&e^{i(\beta_1-\beta_2+\beta_3)}\sin\theta_3\sin\theta_2 \,b\Big]|-\rangle \Big\}\otimes| {-1}\rangle\nonumber\\
&+&e^{i(\beta_1-\alpha_2-\alpha_3)}\cos\theta_3\cos\theta_2 \,b |-\rangle\otimes | {-3}\rangle, 
\label{psi_three}
\end{eqnarray}
where, as before, every introduced magnitude appears in some point of the expression. We have to check $\rho(n,t)$ and $P_{\pm}(t)$ in order to decide which of them are really relevant to the problem:  
\begin{eqnarray}
\rho(3,3)&=&\cos^2\theta_3\cos^2\theta_2\left|a\right|^2,\\
\rho(1,3)&=&\sin^2\theta_3 \left|a\right|^2+
\cos^2\theta_3\sin^2\theta_2 \left|b\right|^2\nonumber\\
&+&\sin 2\theta_3\sin^2\theta_2\, c,\\
\rho(-1,3)&=&\cos^2\theta_3\sin^2\theta_2\left|a\right|^2 +\sin^2\theta_3 \left|b\right|^2\nonumber\\
&-&\sin 2\theta_3\sin^2\theta_2\, c,\\
\rho(-3,3)&=&\cos^2\theta_3\cos^2\theta_2\left|b\right|^2,
\end{eqnarray}
and
\begin{eqnarray}
P_{\pm}(t=3)&=&\frac{1}{2}\left[1\pm\cos 2\theta_3\cos 2\theta_2\left(\left|a\right|^2-\left|b\right|^2\right)\right]\nonumber\\
&\pm&\sin 2\theta_3\sin^2\theta_2\, c,
\end{eqnarray}
with
\begin{equation}
c\equiv{\rm Re}\left[ e^{i(\alpha_1-\alpha_3)}e^{-i(\beta_1-2\beta_2+\beta_3)}b^*a\right],
\label{c_def}
\end{equation}
an interference term. Note that $c$ depends on $\alpha_3$ and $\beta_3$, but also on $\beta_2$, which has reappeared, and observe how $\alpha_2$ is still missing.

The complexity of the explicit expressions for $|\psi\rangle_{t}$ and the derived magnitudes when  $t>3$ is so high that we are not going to reproduce them here. However, as we will show below, Eq.~\eqref{c_def} contains enough clues to understanding the main features of the process. 

\section{Changing phases}
\label{Sec_Changing_phases}

A first conclusion that one can obtain from the previous Section is that $\alpha_t$, $\beta_t$ and $\theta_t$ are significant magnitudes. The importance of a time-dependent $\theta_t$ in the evolution of the quantum walker was beyond any doubt, as this has been the target of many previous studies~\cite{RMM04,AR09a,AR09b,RS14}, so we will fix its value constant $\theta_t=\theta$, and concentrate our attention in the effects in $\hat{U}_t$ of evolving phases~\cite{BNPRS06}.

Another outstanding fact, shown by Eq.~\eqref{c_def}, is that the roles played by $\alpha_t$ and $\beta_t$ are not interchangeable, so we are going to consider them separately. Note that in Ref.~\cite{BNPRS06} the restriction $\alpha_t=-\beta_t$ was imposed. We will discuss some consequences of this constraint later on.

If $\alpha_t$ and $\beta_t$ can be chosen independently, further simplification can be obtained by dropping $\gamma$. This parameter appears in all the previous expressions only through $\varphi$, cf. Eq.~\eqref{phi_def}, 
so we can set $\gamma=0$ hereafter, without any loss of generality 
as long as we do not set $\alpha_1=-\beta_1$.

The fact that $\alpha_t$ and $\beta_t$ affect the evolution of the quantum walker in a distinctive way does not imply, however, that we cannot obtain similar results with both of them. For instance, a source of randomness in any of the two phases will distort the coherence of the evolution of the walker~\cite{AVWW11}, which shifts from a ballistic movement to a diffusive spreading~\cite{BCA03b}. 

\begin{figure}[htbp]
\begin{tabular}{rc}(a)&\includegraphics[width=0.9\columnwidth,keepaspectratio=true]{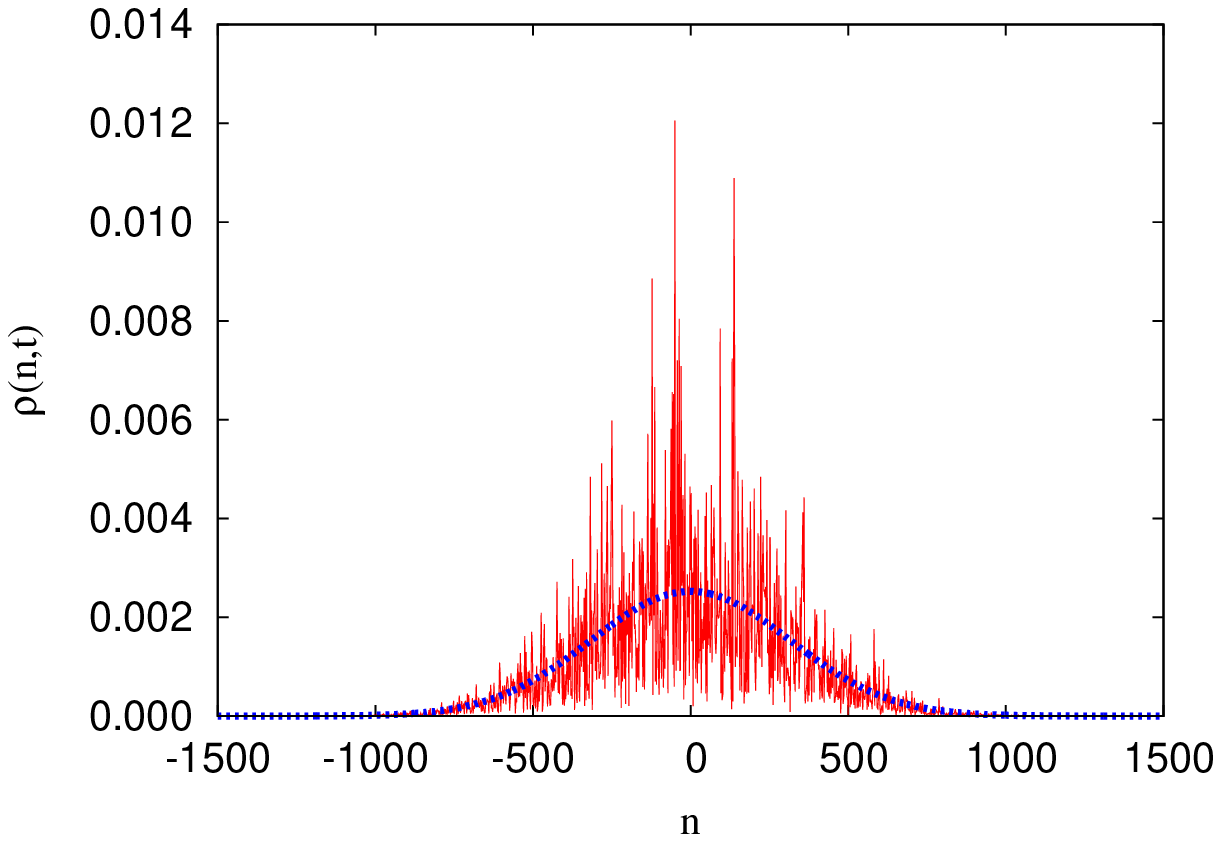}\\(b)&\includegraphics[width=0.9\columnwidth,keepaspectratio=true]{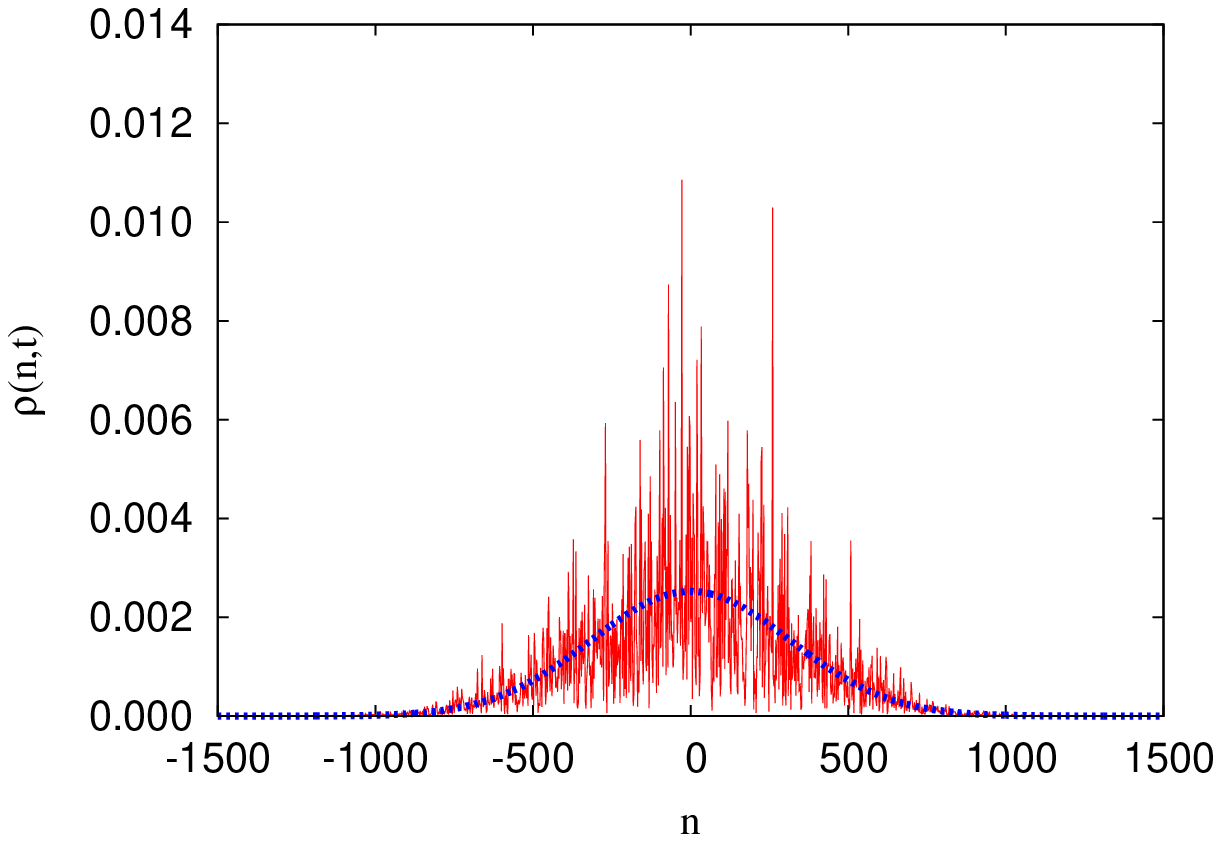}\end{tabular}
\caption{(Color online) 
Probability mass function of the process after $t=100\,000$ time steps.  The red solid line connects the points obtained by direct application of the evolution operator on the initial state when we pick at random: (a) $\alpha_t$; (b) $\beta_t$. Only probabilities corresponding to even values of $n$ are represented, as odd values have probability equal to zero. 
The blue dotted line corresponds to the limiting Gaussian probability density function.} 
\label{Fig_Random}
\end{figure}

We show this in Fig.~\ref{Fig_Random}. In the upper panel we set $\beta_t=0$ and choose $\alpha_t$ at random, whereas the lower one is the outcome of the complementary experiment, we keep $\alpha_t=0$ and change $\beta_t$. The rest of the parameters were fixed to $\theta=\eta=\pi/4$. As we can see, in both cases the quantum walk reaches the classical limit, in the classical way: the PMF presents the archetypal hairy aspect, with sharp spikes above and below the Gaussian bell~\cite{RMM04}. As usual, we have plotted only the non-zero values of PMF, since Eq.~\eqref{evol_t} implies that if the initial state is located at the origin, cf. Eq.~\eqref{psi_zero_gen}, at any later time only odd or even sites are occupied with non-zero probabilities.

Another effect that can be induced with any of the two phases is the breaking of the space symmetry. We can illustrate this easily if we recall the explicit expressions introduced in the previous section. Let us assume again that $\theta=\eta=\pi/4$, and consider that either $\alpha=\pi/2$ and $\beta=0$, or $\alpha=0$ and $\beta=\pi/2$, i.e., $\varphi=\pi/2$. It is well-known that these values lead to a symmetric PMF around the center of the line, whenever the phases are constant. In such a case one has:
\begin{equation*}
\rho(1,1)=\rho(-1,1)=\frac{1}{2},
\end{equation*}
at $t=1$,
\begin{eqnarray*}
\rho(2,2)&=&\rho(-2,2)=\frac{1}{4},\\
\rho(0,2)&=&\frac{1}{2},
\end{eqnarray*}
at $t=2$, but
\begin{eqnarray*}
\rho(3,3)&=&\rho(-3,3)=\frac{1}{8},\\
\rho(1,3)&=&\frac{3}{8}+\frac{c}{2}\\
\rho(-1,3)&=&\frac{3}{8}-\frac{c}{2},
\end{eqnarray*}
with 
\begin{equation*}
c=\frac{1}{2}\sin \left[(\alpha_1-\alpha_3)-(\beta_1-2\beta_2+\beta_3)\right].
\end{equation*}
When $c\neq0$ we depart from a symmetric PMF, and the maximal skewness is attained when 
\begin{equation*}
|c|=\frac{1}{2},
\end{equation*}
that can be obtained, for instance, if set $\alpha_t=\beta_t=0$, for $t>1$. 

Just on the opposite side, we could question if $\alpha_t$ and $\beta_t$ can be tuned so that one recovers the {\it same\/} evolution of a time-independent quantum walker.
Obviously, we are looking for a nontrivial response, and this response cannot stem from the restricted example analyzed in the previous paragraph: $c\neq0$ destroys the symmetry but $c=0$ does not ensures the desired invariance. However, we can re-examine Eq.~\eqref{c_def} in a search for inspiration. Let us focus on the evolution of $\beta_t$, and consider for the moment that $\alpha_t=0$. It is clear that the two choices,
$\beta_{3}=2\beta_{2}-\beta_{1}$ and $\beta_{3}=\beta_{2}=\beta_{1}$, are equivalents, in the sense that both rules lead to the same values of $\rho(n,t)$ and $P_{\pm}(n,t)$, for $t\leq3$. Therefore, Eq.~\eqref{c_def} seems to suggest that a sufficient condition for the invariance is:
\begin{eqnarray}
\alpha_{t}&=&0,\label{alpha_rec_0}\\ 
\beta_{t}&=&2\beta_{t-1}-\beta_{t-2}.\label{beta_rec}
\end{eqnarray}
In the next Section we demonstrate the general validity of this statement, as well as consider the invariance obtained by adjusting $\alpha_t$ alone.

\section{Invariance}
\label{Sec_Invariance}

If we set $\alpha_t=0$, $\hat{U}_t$ reads
\begin{eqnarray}
\hat{U}_t&=& \big[\cos \theta_t |+\rangle  \langle +| +e^{-i\beta_t}\sin \theta_t  |+\rangle  \langle -| \nonumber \\
&+&e^{i\beta_t}\sin \theta_t  |-\rangle  \langle +| -\cos \theta_t   |-\rangle  \langle -|\big]\otimes\hat{I}_P.
\label{U_coin}
\end{eqnarray}
Let us now introduce the wave functions $\psi_{\pm}(n,t)$, the two-dimensional projection of the state of the walker into the position basis:
\begin{eqnarray}
\psi_{+}(n,t)&\equiv& \langle +|  \otimes  \langle n| \psi\rangle_t, \label{Def_Psi_P}\\
\psi_{-}(n,t)&\equiv& \langle -|  \otimes  \langle n| \psi\rangle_t. \label{Def_Psi_M} 
\end{eqnarray}
The evolution operator $\hat{T}_t$ induces the following set of recursive equations on the wave-function components, cf. Eq.~\eqref{evol_t}:
\begin{equation}
\psi_{+}(n,t)=\cos \theta \,\psi_{+}(n-1,t-1)+e^{-i\beta_t}\sin \theta \,\psi_{-}(n-1,t-1),
\label{Rec_P}
\end{equation}
and
\begin{equation}
\psi_{-}(n,t)=e^{i\beta_t}\sin \theta \,\psi_{+}(n+1,t-1)
-\cos \theta\, \psi_{-}(n+1,t-1),
\label{Rec_M}
\end{equation}
which are to be solved under the assumption that the walker is initially at $n=0$, that is, $\psi_{+}(n,0)=\cos \eta\, \delta_{n,0}$, $\psi_{-}(n,0)=\sin \eta\, \delta_{n,0}$, where $\delta_{n,t}$ is the Kronecker delta.

Let us introduce the allied quantities $\psi^{\circ}_{\pm}(n,t)$, the solutions to the time-independent problem, i.e., when $\beta_t=\beta_1$:
\begin{equation}
\psi^{\circ}_{+}(n,t)=\cos \theta \,\psi^{\circ}_{+}(n-1,t-1)+e^{-i\beta_1}\sin \theta \,\psi^{\circ}_{-}(n-1,t-1),
\label{Rec_P0}
\end{equation}
and
\begin{equation}
\psi^{\circ}_{-}(n,t)=e^{i\beta_1}\sin \theta \,\psi^{\circ}_{+}(n+1,t-1)
-\cos \theta\, \psi^{\circ}_{-}(n+1,t-1).
\label{Rec_M0}
\end{equation}
Let us further assume that $\psi_{\pm}(n,t)$ and $\psi^{\circ}_{\pm}(n,t)$ are connected through the following relationships:  
\begin{equation}
\psi_{+}(n,t)=\psi^{\circ}_{+}(n,t)e^{i(n-t)(\beta_2-\beta_1)/2},
\label{psi_psi0_P}
\end{equation}
and
\begin{equation}
\psi_{-}(n,t)=\psi^{\circ}_{-}(n,t)e^{i(n+t)(\beta_2-\beta_1)/2}.
\label{psi_psi0_M}
\end{equation}
Note that Eqs.~\eqref{psi_psi0_P} and~\eqref{psi_psi0_M} satisfy $\psi_{\pm}(0,0)=\psi^{\circ}_{\pm}(0,0)$. Let us consider Eq.~\eref{Rec_P},
\begin{eqnarray*}
\psi_{+}(n,t)&=&\psi^{\circ}_{+}(n,t)e^{i\frac{n-t}{2}(\beta_2-\beta_1)}\\
&=&\cos \theta \,\psi^{\circ}_{+}(n-1,t-1)e^{i\frac{n-t}{2}(\beta_2-\beta_1)}\\
&+&e^{-i\beta_t}\sin \theta \,\psi^{\circ}_{-}(n-1,t-1)e^{i\frac{n+t-2}{2}(\beta_2-\beta_1)},
\end{eqnarray*}
and compare it with Eq.~\eref{Rec_P0}: one must conclude that
\begin{equation}
\beta_t-\frac{n+t-2}{2}(\beta_2-\beta_1)+\frac{n-t}{2}(\beta_2-\beta_1)=\beta_1,
\end{equation}
should hold, that is
\begin{equation}
\beta_{t}=\beta_1+(t-1)\left(\beta_{2}-\beta_{1}\right),
\label{beta_explicit}
\end{equation}
for $t\geq1$. The same conclusion is obtained from Eqs.~\eref{Rec_M},~\eref{Rec_M0} and~\eref{psi_psi0_M}.

From Eq.~\eqref{beta_explicit} it can be easily checked that Eq.~\eqref{beta_rec} is satisfied, and that the recursive law may also be expressed in the following suggesting form:
\begin{equation}
\beta_{t+1}-\beta_{t}=\beta_{t}-\beta_{t-1}.
\end{equation}

Observe how by confirming the validity of Eq.~\eqref{psi_psi0_P} and~\eqref{psi_psi0_M} we have proven that neither $\rho(t,n)$ nor $P_{\pm}(t)$ depend on $\beta_2$, since
\begin{equation}
\left|\psi_{\pm}(n,t)\right|=\left|\psi^{\circ}_{\pm}(n,t)\right|.
\end{equation}
Therefore, we cannot use the detailed knowledge of  $\rho(t,n)$ or $P_{\pm}(t)$ to deduce the right value of $\beta_2$ which is the key to continue with the evolution of the system. The same applies to $M(n,t)$,
\begin{equation}
M(n,t) \equiv\left|\psi_{+}(n,t)\right|^2-\left|\psi_{-}(n,t)\right|^2,
\end{equation}
another interesting magnitude that is connected with the local magnetization of the system in the $z$ direction when the qubit represents the spin of the particle~\cite{SA13}. If we continue with this analogy, $\beta_2$ can only be inferred from the local magnetic properties of the system along the $x$ or $y$ directions.

\begin{figure}[htbp]
\begin{tabular}{rc}
(a)&\includegraphics[width=0.9\columnwidth,keepaspectratio=true]{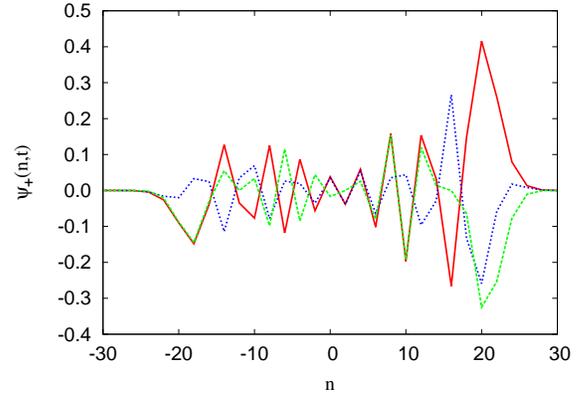}\\(b)&\includegraphics[width=0.9\columnwidth,keepaspectratio=true]{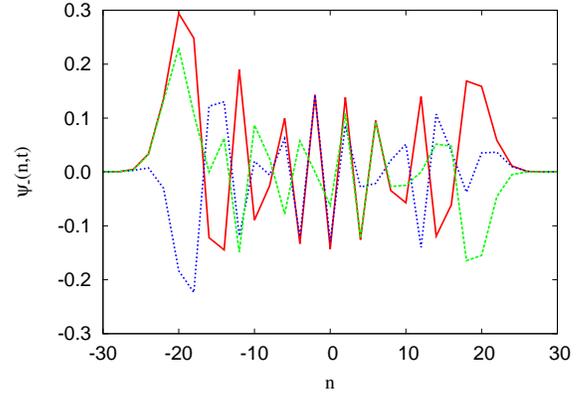}\\(c)&\includegraphics[width=0.9\columnwidth,keepaspectratio=true]{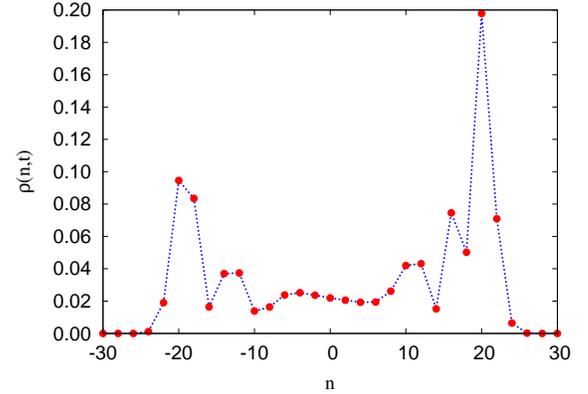}
\end{tabular}
\caption{(Color online) 
Comparison of the wave function after $t=30$ time steps. The red solid lines and dots correspond to a time-homogeneous QW. 
The blue dotted lines show the real parts of the magnitudes associated with a time-dependent QW, 
while the imaginary parts are depicted by green dashed lines.} 
\label{Fig_Sample}
\end{figure}

We illustrate in Fig.~\ref{Fig_Sample} the invariance of $\rho(t,n)$ in spite of the time- and site-inhomogeneous phase shifts that Eq.~\eqref{beta_explicit} introduces in the wave functions, cf. Eqs.~\eqref{psi_psi0_P} and~\eqref{psi_psi0_M}. Here we have set $\theta=\pi/4$,  $\eta=\pi/3$,  $\alpha_t=0$, $\beta_1=0$ and $\beta_2=\pi/7$. With this choice, $\psi^{\circ}_{\pm}(n,t)$ are real functions, whereas $\psi_{\pm}(n,t)$ exhibit a complex, correlated behavior: their relative contribution to the PMF changes abruptly between neighbor locations. 

We can draw a complementary picture that may help in the understanding the evolution of $\hat{U}_t$ when $\beta_t$ follows Eq.~\eqref{beta_explicit}, through a geometrical analogy, a representation that is very similar to the Bloch sphere~\cite{AR09b}. Let us introduce $\boldsymbol{u}_t$, a time-dependent, unit-length vector in $\RR^3$. Let us denote by $\theta$ and $\beta_t$ its polar and azimuthal spherical coordinates, respectively. Then, we can recover the coin operator $\hat{U}_t$ through the scalar projection of the Pauli {\it vector\/} $\hat{\boldsymbol{\sigma}}$, with Cartesian components 
\begin{eqnarray*}
\hat{\sigma}_x&\equiv&  |+\rangle  \langle -| + |-\rangle  \langle +|,\\
\hat{\sigma}_y&\equiv& -i |+\rangle  \langle -| + i |-\rangle  \langle +|, \mbox{ and}\\
\hat{\sigma}_z&\equiv& |+\rangle  \langle +|  -  |-\rangle  \langle -|,
\end{eqnarray*}
onto the $\boldsymbol{u}_t$ direction,  i.e.,
\begin{equation}
\hat{U}_t\equiv \left(\boldsymbol{u}_t\cdot\hat{\boldsymbol{\sigma}}\right)\otimes\hat{I}_P.
\end{equation}
The evolution of $\boldsymbol{u}_t$ is a step-like precession around the north pole. Observe how, unlike the example shown in Fig.~\ref{Fig_Sample}, $(\beta_2-\beta_1)/\pi$ does not need to be a rational fraction and, therefore, the precession of $\boldsymbol{u}_t$ is not a periodic phenomenon in general. The absence of periodicity implies that the succession defined by the ending points of the vector $\boldsymbol{u}_t$ constitutes an everywhere-dense subgroup of the corresponding parallel of latitude on the (Bloch) sphere, and thus the unconditional probability of choosing a particular value for $\beta_t$ is uniformly distributed in the limit.

This feature is not shared by the second path to invariance, i.e.,
\begin{eqnarray}
\alpha_{t}&=&\alpha_{t-2},\label{alpha_rec}\\ 
\beta_{t}&=&\beta_{t-1}.\label{beta_rec_0}
\end{eqnarray}
As one can easily see, Eq.~\eqref{alpha_rec} implies that $\alpha_t$ can take no more than two different values, $\alpha_1$ and $\alpha_2$, and that these two phases must be chosen alternately. The formal proof follows the same steps as in the previous case, so we are going to give a simple sketch of it. We have to assume that the solution to our problem reads
\begin{equation}
\psi_{+}(n,t)=\psi^{\circ}_{+}(n,t)e^{i n(\alpha_2-\alpha_1)/2},
\end{equation}
and
\begin{equation}
\psi_{-}(n,t)=\psi^{\circ}_{-}(n,t)e^{i n (\alpha_2-\alpha_1)/2},
\end{equation}
when $t$ is even, and
\begin{equation}
\psi_{+}(n,t)=\psi^{\circ}_{+}(n,t)e^{i(n-1)(\alpha_2-\alpha_1)/2},
\end{equation}
and
\begin{equation}
\psi_{-}(n,t)=\psi^{\circ}_{-}(n,t)e^{i(n+1)(\alpha_2-\alpha_1)/2},
\end{equation}
when $t$ is odd. The introduction of these expressions in the set of recursive formulas induced by Eq.~\eqref{evol_t},
\begin{equation*}
\psi_{+}(n,t)=e^{i\alpha_t}\cos \theta \,\psi_{+}(n-1,t-1)+e\sin \theta \,\psi_{-}(n-1,t-1),
\end{equation*}
and
\begin{equation*}
\psi_{-}(n,t)=\sin \theta \,\psi_{+}(n+1,t-1)
-e^{-i\alpha_t}\cos \theta\, \psi_{-}(n+1,t-1),
\end{equation*}
leads to the periodic alternation of $\alpha_t$ between $\alpha_1$ and $\alpha_2$, which is still arbitrary. Even so, this case is not as rich as the previous one.

Finally, note that we have not ruled out the possibility that this kind of nontrivial invariance can be generated by means of recursive laws where $\alpha_t$ and $\beta_t$ collaborate together. However, we can show that the phase locking 
$\alpha_t=-\beta_t$ is not a valid candidate. The key point for this effect is that value of the phase factors at time $t=2$ can be set with independence of the phase factors at time $t=1$. This implies, in particular, that condition $c=0$ must be identically satisfied for any choice of $\beta_2$. When $\alpha_t=-\beta_t$, the expression for $c$ reads, cf. Eq.~\eqref{c_def},
\begin{equation*}
c={\rm Re}\left[ e^{i2(\beta_2-\beta_1)}b^*a\right],
\end{equation*}
and then $\beta_2$ is tied to $\beta_1$.

\section{Conclusion}
\label{Sec_Conclusion}

In this paper we have analyzed some properties of a discrete-time quantum walk on the line when the coin operator depends on the time variable. 

In particular, we have focused our interest on the effects that the time dependence of the two phase factors can cause on the behavior of the walker. In the first place, we have shown how random changes in any of these magnitudes lead to the recovery of the classical limit. But decoherence is not the immediate consequence of temporal inhomogeneity in the process. On the contrary, we have found a way to mimic the motion of any ordinary quantum walk by means of a sequence of well-chosen, time-dependent coin operators.

A remarkable property of this invariance is that the way in which the {\it replicating\/} sequence must be designed is precise but not unique. It depends on two free parameter whose values cannot be assessed through the mere inspection of the position of the walker.

The information that is hidden in these two magnitudes must be recovered by means of specific inspection of the local quantum properties coined walker. The implications of this fact will be the subject of future research.

\acknowledgments
The author acknowledges partial support from the Spanish Ministerio de Econom\'{\i}a y Competitividad under Contract No. FIS2013-47532-C3-2-P, and from Generalitat de Catalunya, Contract No. 2014SGR608.


\end{document}